\documentclass[letterpaper, 10 pt, conference]{ieeeconf}  
\IEEEoverridecommandlockouts                              
\usepackage{graphicx} 
\usepackage{amsmath} 
\usepackage{amssymb}  
\usepackage{bm}
\usepackage{amsfonts,textcomp,empheq} 
\usepackage{subfigure}
\usepackage{booktabs}
\usepackage{xspace}
\usepackage[ruled,noend,norelsize]{algorithm2e}
\usepackage{color}
\usepackage[bookmarks=false,colorlinks]{hyperref}
\AtBeginDocument{%
  \hypersetup{
    linkcolor=blue,   
    citecolor=red,
    urlcolor=magenta}}

\newcommand{\Ampl}{\textsc{Ampl}\xspace}
\newcommand{\Cplex}{\textsc{Cplex}\xspace}
\newcommand{\Ipopt}{\textsc{Ipopt}\xspace}

\newcommand{\pit}{\bm{p}_i(t)}
\newcommand{\pjt}{\bm{p}_j(t)}
\newcommand{\prt}{\bm{p}_{ij}(t)}
\newcommand{\vr}{\bm{v}_{ij}}
\newcommand{\vrx}{v_{x}}
\newcommand{\vry}{v_{y}}
\newcommand{\dix}{\delta_{i,x}}
\newcommand{\diy}{\delta_{i,y}}
\newcommand{\tdix}{\tilde{\delta}_{i,x}}
\newcommand{\tdiy}{\tilde{\delta}_{i,y}}

\newcommand{\vix}{v_{i,x}}
\newcommand{\viy}{v_{i,y}}
\newcommand{\vjx}{v_{j,x}}
\newcommand{\vjy}{v_{j,y}}
\newcommand{\pz}{\bm{\widehat{p}}_{ij}}
\newcommand{\pzx}{\widehat{x}}
\newcommand{\pzy}{\widehat{y}}

\newcommand{\viz}{\widehat{v}_{i}}
\newcommand{\vjz}{\widehat{v}_{j}}

\newcommand{\qi}{q_{i}}
\newcommand{\qj}{q_{j}}
\newcommand{\ti}{\theta_{i}}
\newcommand{\tj}{\theta_{j}}
\newcommand{\qmax}{\overline{q}}
\newcommand{\qmin}{\underline{q}}
\newcommand{\tmax}{\overline{\theta}}
\newcommand{\tmin}{\underline{\theta}}
\newcommand{\tiz}{\widehat{\theta}_{i}}
\newcommand{\tjz}{\widehat{\theta}_{j}}
\newcommand{\tm}{t^m_{ij}}
\newcommand{\g}{g}
\newcommand{\f}{f_{ij}(t)}
\newcommand{\ftm}{f_{ij}(\tm)}
\newcommand{\fp}{f'_{ij}(t)}

\newcommand{\z}{z}

\newcommand{\vi}{\bm{v}_i}
\newcommand{\ii}{\mathrm{i}}

\newtheorem{model}{Model}
\DeclareMathOperator{\atantwo}{arctan2}

\title{\LARGE \bf
Complex Number Formulation and Convex Relaxations\\ 
for Aircraft Conflict Resolution
}

\author{David Rey and Hassan Hijazi
\thanks{D. Rey is with School of Civil and Environmental Engineering, UNSW Sydney,
        2052, NSW, Australia
        {\tt\small d.rey@unsw.edu.au}}%
\thanks{H. Hijazi is with Los Alamos National Laboratory,
        Los Alamos, 87544, NM, USA.
        {\tt\small hlh@lanl.gov}}%
}

\begin{document}

\maketitle
\thispagestyle{empty}
\pagestyle{empty}

\begin{abstract}
We present a novel complex number formulation along with tight convex relaxations for the aircraft conflict resolution problem. Our approach combines both speed and heading control and provides global optimality guarantees despite non-convexities in the feasible region. We present a new characterization of the conflict separation condition in the form of disjunctive linear constraints. Using our approach, we are able to close a number of open instances and reduce computational time by up to two orders of magnitude on standard instances.
\end{abstract}

\section{INTRODUCTION}

Safety plays a critical role in ATM due to the high stakes involved in aircraft operations. The safety of flights is ensured by Air Traffic Control (ATC) services which are in charge of monitoring aircraft trajectories and maintaining minimum separation distances between aircraft \cite{Nol10}. Current air separation standards issued by the International Civil Aviation Organization (ICAO) require that aircraft be separated by at least 5 NM\footnote{Nautical Mile---1 NM = 1852 m.} horizontally and 1000 ft vertically \cite{ICA96} and two aircraft violating these rules are said to be in \emph{conflict}. In this paper, we present a novel Conflict Detection and Resolution (CD\&R) algorithm based on speed and heading control for en-route traffic. We focus on the horizontal aircraft conflict resolution problem \emph{i.e.} 2-dimensional plane, since altitude change maneuvers can be easily modeled using discrete decision variables. \newline

The aircraft conflict resolution problem is traditionally represented as an optimization problem in which the objective is to find conflict-free trajectories for all aircraft flying in a given region of airspace. A comprehensive review of the literature on CD\&R algorithms up to the 21st century can be found in \cite{KY00}. Since we propose a global optimization approach for the aircraft conflict resolution problem, we next focus on reviewing the literature on exact methods. 

The first exact approaches for conflict resolution are due to \cite{RH02} and \cite{PFB02}. In \cite{RH02}, the authors propose a Mixed-Integer Linear Program (MILP) to find conflict-free aircraft trajectories in the horizontal space. Aircraft dynamics are approximated and separation constraints are verified at discrete time steps. An optimal control formulation with speed and heading maneuvers is proposed and solved on instances with up to 4 aircraft. In \cite{PFB02}, two horizontal conflict resolution problems are solved: a first problem is solved with speed control only and a second with heading control. In \cite{VSSC09}, the authors present a MILP for speed and altitude control based on a disjunctive linear separation constraint. This separation condition is also used in \cite{RRFF15} where the authors introduce linear upper bounds when speed control is the only separation maneuver. \cite{Ome15} proposes a space-discretized MILP formulation involving speed and heading controls where aircraft recover a parallel trajectory. 

Several papers have built on the separation constraints introduced in \cite{PFB02}. In \cite{AEM11}, the problem is represented as a MILP and improvements on the original formulation are proposed. In \cite{AEM14}, a Mixed-Integer Non-Linear Program (MINLP) is proposed to solve the horizontal aircraft collision avoidance problem. The separation constraints are expressed using trigonometric functions to represent heading variations. Recently, in \cite{AEM16}, the authors proposed an exact non-convex MINLP approach combining speed, heading and altitude controls. The authors also consider a 2D variant of their problem in which only horizontal maneuvers (speed and heading control) are allowed which coincides with the problem we are addressing in this paper. Let us emphasise that, in \cite{AEM11, AEM14, AEM16}, the authors point out the existence of past conflicts in their formulations without providing a comprehensive method to handle these. In particular, if heading control is allowed, initially diverging aircraft (before optimization) can eventually converge (after optimization). Alternative separation conditions derived from aircraft pairs time of minimal distance were recently proposed in \cite{CR17,CO16} and formulated using MINLPs. 

This review highlights that exact approaches are penalized by the non-convexity of trigonometric functions involved in the separation conditions. In turn, discretised approaches either use upper bounds on aircraft minimal crossing times to guarantee conflict-free trajectories or consider only a finite number of alternative trajectories, thus potentially ignoring conflict-free solutions with better objective function values. In this paper, we present a new formulation for aircraft separation based on a complex number representation of velocity control. We then introduce convex relaxations in the form of Mixed-Integer Quadratic Programs (MIQPs) and a Mixed-Integer Quadratically-Constrained Programs (MIQCPs). We show that these convex relaxations are likely to produce global optimal solutions, \emph{i.e.}, the relaxations are usually tight. Numerical results highlight the efficiency of the proposed approach when compared to state-of-the-art methods on classical benchmark problems.

\section{AIRCRAFT SEPARATION CONDITION}

Let us consider a set $A$ of aircraft in a given air sector, all at the same flight level. Let $\pit = [x_i(t), y_i(t)]^\intercal$ be the vector representing the position of flight $i$ at time $t$. The relative position of aircraft $i$ and $j$ at time $t$ can be represented as $\prt = \pit - \pjt$. Let $d$ be the horizontal separation norm, the two aircraft are separated \mbox{if and only if}:
\begin{equation}
\|\prt \| \geqslant d, \forall t \geqslant 0
\label{eq:sep}
\end{equation}

Let $\vr = [v_{ij,x}, v_{ij,y}]^\intercal$ be the relative velocity vector of $i$ and $j$, \emph{i.e.} $v_{ij,x} = \vix - \vjx$ and $v_{ij,y} = \viy - \vjy$, and let $\pz = [\pzx_{ij}, \pzy_{ij}]^\intercal$ be their relative initial positions. Assuming that uniform motion laws apply, $\prt$ can be expressed as: $\prt = \pz + \vr t$.

For each aircraft $i \in A$, we denote $\viz$ its initial speed and $\tiz$ its initial heading. Let $\qi$ be the speed variation rate ($\qi = 1$ means no deviation) and let $\tj$ be the heading deviation angle ($\ti = 0$ means no deviation): $\qi$ and $\ti$ are the main speed and heading control variables for $i \in A$, respectively. Aircraft velocity components are $\vix = \qi\viz \cos (\ti + \tiz)$ and $\viy = \qi\viz \sin (\ti + \tiz)$. Aircraft relative velocity vector components can then be written as
\begin{align}
v_{ij,x} &= \qi\viz \cos (\ti + \tiz) - \qj\vjz \cos (\tj + \tjz) \label{eq:vrx}\\
v_{ij,y} &= \qi\viz \sin (\ti + \tiz) - \qj\vjz \sin (\tj + \tjz) \label{eq:vry}
\end{align}


Modeling aircraft separation can be achieved by determining the time of minimum separation based on aircraft relative motion \cite{Caf14,CR17}. Squaring Equation \eqref{eq:sep} we obtain the separation condition
\begin{equation}
\f \equiv \|\vr\|^2 t^2 + 2\pz \cdot \vr t + \|\pz\|^2 - d^2 \geqslant 0
\label{eq:sep2}
\end{equation}

where $\cdot$ is the inner product in the Euclidean space. From Equation \eqref{eq:sep2}, $\f$ is a 2nd order convex polynomial in $t$. Let $\tm$ be the time at which $\f$ is minimal
\begin{equation}
\fp = 0 \quad \Leftrightarrow \quad \tm = \frac{-\pz \cdot \vr}{\|\vr\|^2}
\label{eq:tmin}
\end{equation}
Note that the sign of the inner product $\pz \cdot \vr$ indicates aircraft convergence/divergence, formally we can state that, $\pz \cdot \vr >0 \Leftrightarrow \tm < 0$ which indicates that aircraft $i$ and $j$ are diverging. Modeling aircraft divergence is critical if heading control maneuvers are allowed since a pair of initially diverging aircraft (before optimization) may converge (after optimization).

Substituting $\tm$ in $\f$, the separation condition \eqref{eq:sep2} can be simplified to $\ftm \geqslant 0$, which does not depend on $t$ anymore. Furthermore, multiplying both sides by $\|\vr\|^2$ gives the following separation condition
\begin{align}
g_{ij}(\vr) &= \|\vr\|^2\ftm \nonumber \\ 
   &= \|\vr\|^2(\|\pz\|^2 - d^2) - (\pz \cdot \vr)^2 \geqslant 0
\label{eq:sep3}
\end{align}

For clarity of presentation, we will drop the ${ij}$ subscript in the remainder of this section.
In scalar form, the separation condition \eqref{eq:sep3} can be written as a function of aircraft relative velocity:
{\small
\begin{align}
\g (\vrx,\vry) &= \vrx^2 (\pzy^2 - d^2) + \vry^2 (\pzx^2 - d^2) - \vrx \vry (2\pzx \pzy) \ge 0
\label{eq:sep4}
\end{align}
}
The separation constraint \eqref{eq:sep3} provides a sufficient condition for aircraft separation but ignores the temporal dimension of the problem as it cannot differentiate between past and future conflicts. In particular, diverging aircraft do not have to satisfy these constraints as they incur no risk of future conflicts (assuming they are initially separated).  

Note that the function $\g$ on the left hand side of \eqref{eq:sep4} is a two-dimensional quadratic function. Thus, the feasible region corresponding to the equation $\g (\vrx,\vry) = 0$ can be described by two linear equations. These equations are characterised by solving the equation $\g (\vrx,\vry) = 0$ treating $\vry$ or $\vrx$ as a constant. The discriminants of the resulting uni-dimensional quadratic functions are 
\begin{empheq}[left=\empheqlbrace~, right={}, outerbox={}]{align}
&\Delta_{\vrx} = 4d^2\vry^2 (\pzx^2 + \pzy^2 - d^2)\label{eq:delta_x}\\
&\Delta_{\vry} = 4d^2\vrx^2 (\pzx^2 + \pzy^2 - d^2)\label{eq:delta_y}
\end{empheq}
Hence the equation admits real roots if $\pzx^2 + \pzy^2 - d^2 \geqslant 0$, which is always true since $i$ and $j$ are assumed to be initially separated. Given the discriminants defined in \eqref{eq:delta_x} and \eqref{eq:delta_y}, points satisfying $\g (\vrx,\vry) = 0$ must satisfy the following set of linear equations,
{
\small
\begin{empheq}[left=\empheqlbrace~, right={}, outerbox={}]{align}
&(\pzy^2 - d^2)\vrx -  \left(\pzx \pzy + d \sqrt{\pzx^2 + \pzy^2 - d^2}\right)\vry = 0 \label{eq:lin1}\\
&(\pzy^2 - d^2)\vrx -  \left(\pzx \pzy - d \sqrt{\pzx^2 + \pzy^2 - d^2}\right)\vry = 0\label{eq:lin2}\\
&(\pzx^2 - d^2)\vry -  \left(\pzx \pzy - d \sqrt{\pzx^2 + \pzy^2 - d^2}\right)\vrx = 0 \label{eq:lin3}\\
&(\pzx^2 - d^2)\vry -  \left(\pzx \pzy + d \sqrt{\pzx^2 + \pzy^2 - d^2}\right)\vrx = 0\label{eq:lin4}
\end{empheq}
}

Equations \eqref{eq:lin1}-\eqref{eq:lin4} define two lines in the plane $(\vrx,\vry)$ and the sign of $\g (\vrx,\vry)$ depends on these linear equations. Consider the plane equation 
\begin{equation}
\vrx \pzx + \vry \pzy = 0
\label{eq:P}
\tag{P}
\end{equation}
induced by the dot product $\pz \cdot \vr$ and indicating convergence/divergence. This plane splits the space $(\vrx,\vry)$ in two half-spaces, one of which represents diverging trajectories. Any point in this half-space corresponds to diverging trajectories and thus is feasible. The remaining half-space can be split into two symmetric sub-spaces using the plane normal to $\vrx \pzx + \vry \pzy = 0$, defined as,
\begin{equation}
\vry \pzx - \vrx \pzy = 0.
\label{eq:N}\tag{N}
\end{equation}

\begin{figure}[h!]
\centering
\mbox{\subfigure[\scriptsize Side-view of the feasible region]{\includegraphics[width=4cm]{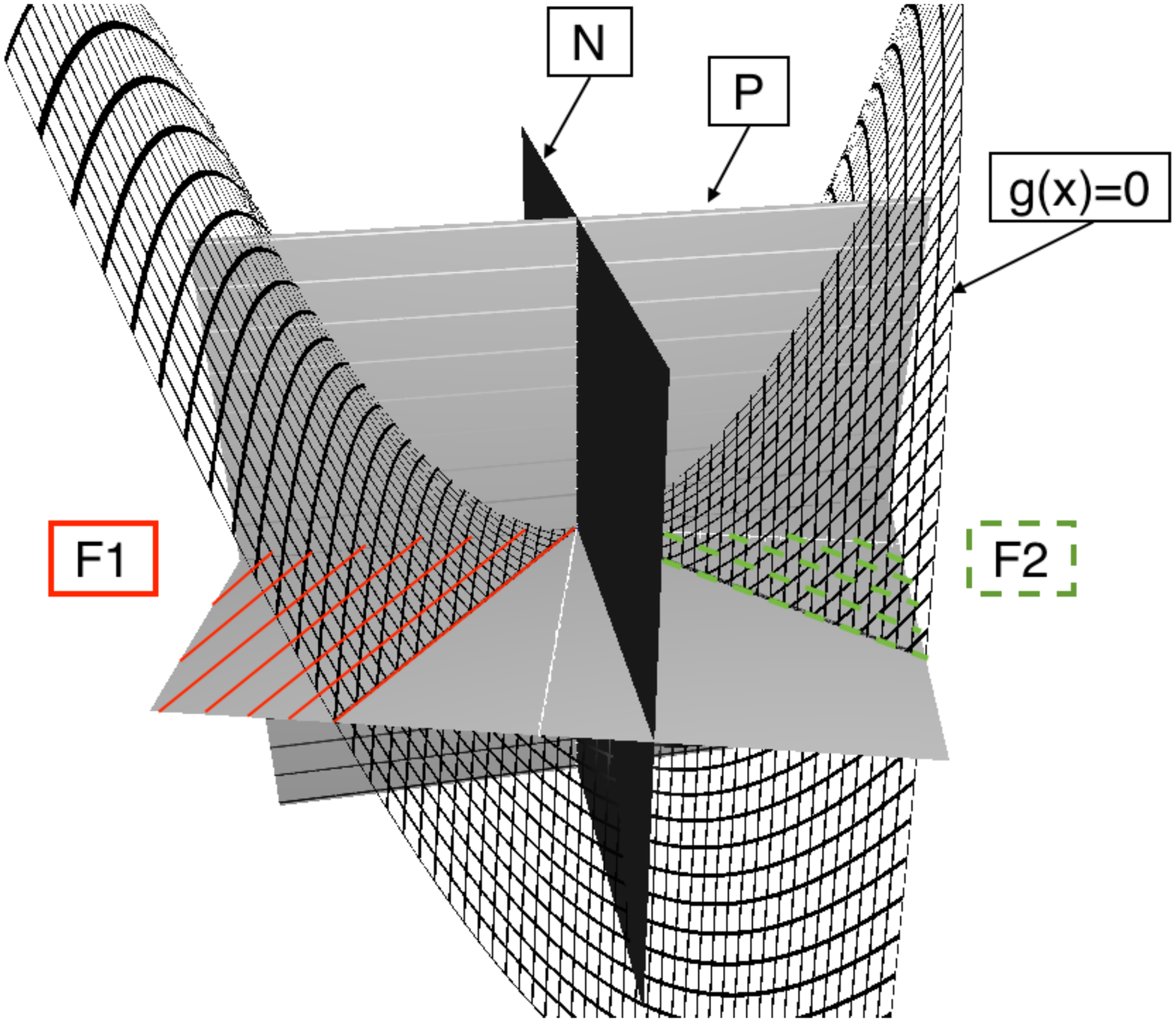}}
\subfigure[\scriptsize Upper-view of the feasible region]{\includegraphics[width=4cm]{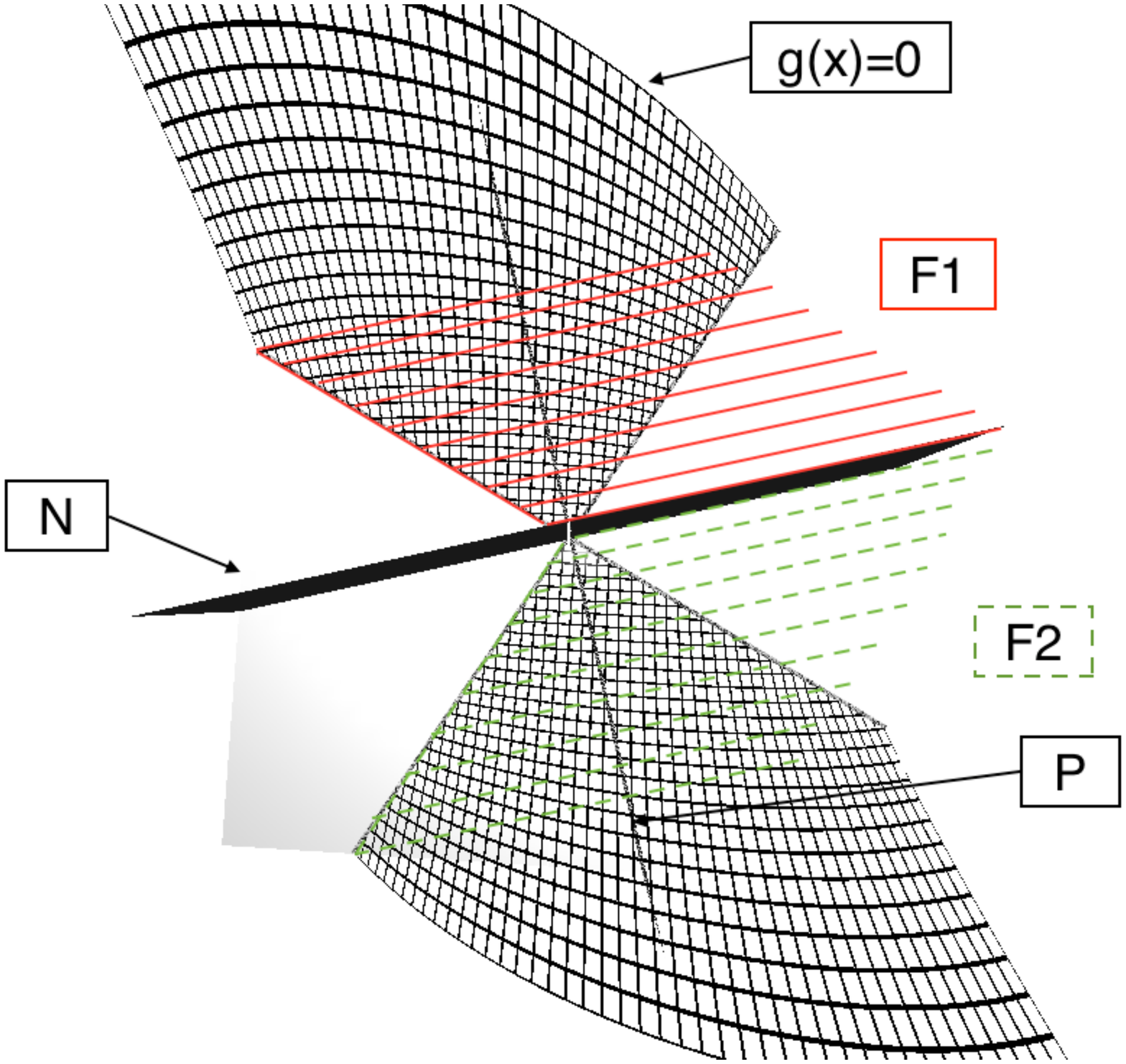} } } 
\caption{Different angles on the feasible region in the plane $(\vrx,\vry)$. Note that points behind the hyperplane (\ref{eq:P}) are feasible (past conflicts) and (\ref{eq:N}) splits the feasible region into two symmetric convex sub-regions denoted $F1$ and $F2$.\label{fig:sep}} 
\end{figure}

Let us emphasise that the feasible region defined by the separation constraint \eqref{eq:sep4} can be reduced to two on/off linear inequalities based on \eqref{eq:lin1}-\eqref{eq:lin4}. Depending on the sign of the constants $\pzx$ and $\pzy$, \eqref{eq:sep4} can only be satisfied on one side of the lines defined by the system of linear equations. An Example of the non-convex region defined by the inequality $\g (\vrx,\vry) \ge 0 $ is depicted in \mbox{Figure \ref{fig:sep}}.

Given a binary variable $z \in \{0,1\}$, let us consider the following disjunction,
$$ \{z = 1,\vry \pzx - \vrx \pzy  \leqslant 0\} \vee \{z = 0,\vry \pzx - \vrx \pzy  \geqslant 0\}.$$
This disjunction models the crossing order of aircraft at the intersection point of their trajectories. Given the disjunction above, the feasible region can be split into two symmetrical polyhedra defined by the lines corresponding to the roots of \eqref{eq:sep4}. We next present a new formulation to link relative velocity variables to aircraft control variables. 

%
%

\section{COMPLEX NUMBER FORMULATION}

Aircraft motion can be represented by the vector $\vi = [\vix, \viy]^\intercal$ where $\vix = \qi\viz \cos (\ti + \tiz)$ and $\viy = \qi\viz \sin (\ti + \tiz)$. We propose to isolate the decision variables $\qi$ and $\ti$ using trigonometric identities:
\begin{align*}
\vix &= \qi \viz \cos(\ti) \cos(\tiz) - \qi \viz \sin(\ti) \sin(\tiz) \\
\viy &= \qi \viz \sin(\ti) \cos(\tiz) + \qi \viz \cos(\ti) \sin(\tiz)
\end{align*}

This representation admits a natural formulation where the control actions are represented as a complex number:
$$V_i = \qi (\cos(\ti) + \ii\sin(\ti))$$
In rectangular form, let $\dix=\Re (V_i)$ and $\diy=\Im (V_i)$ respectively represent the real and imaginary parts of $V_i$, \emph{i.e.},
\begin{equation*}
V_i = \dix+ \ii\diy,~\text{where } \dix = \qi \cos(\ti),~\diy = \qi \sin(\ti). 
\end{equation*}

The magnitude of $V_i$ is then $|V_i| = \sqrt{\dix^2 + \diy^2} = \qi$ and its argument $\arg(V_i) = \atantwo(\diy,\dix) = \ti$. This approach is inspired by complex number formulations for the optimal power flow problem in power systems \cite{Hij_MPC_16,Coff_16}.

A common objective function for aircraft conflict resolution is to minimize the deviation with respect to initial trajectories \cite{PFB02,CO16,AEM16}. This can be achieved by minimizing the norm of both $(1-\qi)$ and $\ti$. Observe that $\diy^2 + (1- \dix)^2 = \qi^2 -2\qi \cos(\ti) + 1$ which is minimal when $\ti = 0 $ and $\qi = 1$. Hence we propose to minimize the objective function: $\sum_{i \in A} \diy^2 + (1- \dix)^2$.

For each $i \in A$, let $0< \qmin < \qmax$ be bounds on $\qi$ and let $\tmin < \tmax$ be bounds on $\ti$. We assume that $\tmin > -\pi/2$ and $\tmax < \pi/2$. This is reasonable since aircraft heading control range is typically limited to $\pm \pi/6$ due to aircraft dynamics and passenger comfort constraints. This implies bounds on $\dix$ and $\diy$:
\begin{align}
\qmin\cos(\max\{|\tmin|,|\tmax|\}) &\leqslant \dix \leqslant \qmax \\
\qmax\sin(\tmin) &\leqslant \diy \leqslant \qmax\sin(\tmax)
\end{align}

Further, observe that $\diy/\dix = \tan(\ti)$ which is smooth between $-\pi/2$ and $\pi/2$ ($\dix > 0$). Hence, the traditional constraints on aircraft control variables $\qi$ and $\ti$ can be expressed in the complex number space $(\dix, \diy)$ as follows:
\begin{align}
\qmin \leqslant \qi \leqslant \qmax &\Leftrightarrow \qmin^2 \leqslant \dix^2 + \diy^2 \leqslant \qmax^2 \label{eq:sbounds}\\
\tmin \leqslant \ti \leqslant \tmax &\Leftrightarrow \dix\tan(\tmin) \leqslant \diy \leqslant \dix\tan(\tmax) \label{eq:hbounds}
\end{align}

The aircraft conflict resolution problem with speed and heading controls is summarized in Model \ref{mod:cnf}, hereby referred to as the Complex Number formulation. Indicator constraints are used to formulate the disjunction therein: depending on the implementation framework, these can be directly passed to the solver or a convex hull formulation can be used based on the methods presented in \cite{Hij_10,Hij_COA_12,Hij_ANU_14}. \\

\begin{model}[Complex Number Formulation]
\label{mod:cnf}
\small
\begin{align*}
&\text{minimize} \sum_{i \in A} \Im (V_i)^2 + (1- \Re (V_i))^2 && \\
&\text{subject to}  && \nonumber \\
&V_{ij} =  V_i\widehat{V}_i - V_j\widehat{V}_j&& \forall (i,j) \in P \nonumber \\
&\Im\left(V_{ij}\widehat{P}_{ij}^*\right)  \leqslant 0 \text{ if } \z=1 &&\forall (i,j) \in P\\
&\Im\left(V_{ij}\widehat{P}_{ij}^*\right) \geqslant 0 \text{ if } \z=0 &&\forall (i,j) \in P\\
&\Im\left(V_{ij}\widehat{L}_{ij}^*\right)  \leqslant 0 \text{ if } \z=1 &&\forall (i,j) \in P\\
&\Im\left(V_{ij}\widehat{U}_{ij}^*\right) \geqslant 0 \text{ if } \z=0 &&\forall (i,j) \in P\\
&\qmin^2 \le |V_i|^2 \leqslant \qmax^2  &&\forall i \in A \\
&\tmin \leqslant \arg(V_i) \leqslant \tmax &&\forall i \in A \\
&V_i,V_{ij} \in \mathbb{C}, z_{ij} \in \{0,1\} &&\forall (i,j) \in P \\
\end{align*}
\end{model}
\normalsize

where $\widehat{V}_{i}= \viz \left(\cos(\tiz) + \ii \sin(\tiz)\right)$, $\widehat{P}_{ij}^*= \pzx_{ij} - \ii\pzy_{ij}$, $\widehat{L}_{ij}^* = \alpha_{ij}^l - \ii\beta_{ij}^l$ and $\widehat{U}_{ij}^* = \alpha_{ij}^u - \ii\beta_{ij}^u$. Note that coefficients $\alpha_{ij}^l$, $\beta_{ij}^l$ and $\alpha_{ij}^u$, $\beta_{ij}^u$ can be preprocessed based on the sign of $\pzx_{ij}$ and $\pzy_{ij}$. For implementation details, a real-number  extension of this model can be found under: \small\url{https://github.com/ReyHijazi/Conflict_Resolution}\normalsize. This formulation is non-convex due to the concave quadratic constraints involved in the left inequality of \eqref{eq:sbounds} and the disjunction modeled by the binary variable $z_{ij}$. We next present convex relaxations for this model.

\section{CONVEX RELAXATIONS AND SOLUTION ALGORITHM}

Non-convexity in the above formulation can be tackled by deriving the convex hull of \eqref{eq:sbounds} as described in \cite{Hij_MPC_16}. Let $\tdix \geqslant 0$ and $\tdiy \geqslant 0$ be variables defined for each $i \in A$ as:
\normalsize
\begin{align}
\qmin^2 &\leqslant \tdix + \tdiy \label{eq:rlmod1}\\
\tdix &\leqslant (1 + \qmin \cos(\max\{|\tmin|,|\tmax|\}))\dix - \qmin \cos(\max\{|\tmin|,|\tmax|\}) \label{eq:rlmod2} \\
\tdiy &\leqslant \qmax (\sin(\tmin) + \sin(\tmax)) \diy - \qmax^2 \sin(\tmin) \sin(\tmax) \label{eq:rlmod3}
\end{align}
\normalsize

Constraints \eqref{eq:rlmod1} set a relaxed lower bound on aircraft speed control while Constraints \eqref{eq:rlmod2} and \eqref{eq:rlmod3} link variables $\tdix$ and $\tdiy$ to convex envelopes of \eqref{eq:sbounds}. Substituting the lower bound on $\dix^2 + \diy^2$ in \eqref{eq:sbounds} by Constraints \eqref{eq:rlmod1}-\eqref{eq:rlmod3} results in a relaxed Mixed-Integer Quadratically Constrained Program (MIQCP) that can be solved by commercial optimization software such as \Cplex \cite{CPLEX09}---we hereby refer to this relaxation as LB-MIQCP. 

The complex number formulation can be further relaxed by entirely omitting Constraints \eqref{eq:sbounds}. While this relaxation ignores aircraft speed control bounds, the resulting formulation is a Mixed-Integer Quadratic Program (MIQP) for which efficient and scalable algorithms are implemented in optimization software---we hereby refer to this relaxation as LB-MIQP. Observe that LB-MIQP is also a relaxation of LB-MIQCP. Formally, let $OPT$ denote the optimal objective value of the complex number formulation and $LB_{MIQP}$ and $LB_{MIQCP}$ be the optimal objective values of LB-MIQP and LB-MIQCP, respectively. The following holds: $LB_{MIQP} \leqslant LB_{MIQCP} \leqslant OPT$. Given the objective function, it is expected that both relaxations LB-MIQP and LB-MIQCP often provide solutions that do not violate aircraft speed control bounds. This is due to the objective function in Model \ref{mod:cnf} aiming at minimizing the deviation to aircraft initial trajectories thus driving $\qi$ away from their bounds. \\

We use the convex relaxations presented above to solve the horizontal aircraft conflict resolution problem. We first solve LB-MIQP and check if the optimal speed vector $\bm{q}^\star$ violates aircraft speed bounds, \emph{i.e.} for each aircraft we check if constraints \eqref{eq:sbounds} is satisfied. If the solution is bound-violating, we then solve LB-MIQCP and check if the newly obtained $\bm{q}^\star$ violates the lower bound in \eqref{eq:sbounds}. If the solution is still bound-violating, we introduce a heuristic to efficiently determine a feasible solution: we fix the binary variable vector $\bm{z}^\star$ and solve Model \ref{mod:cnf} using an interior point method. Note that the Non-Linear Program (NLP) solved in this last step contains only continuous variables and thus provides an upper bound on $OPT$---we hereby refer to this problem as UB-NLP. This solution algorithm is summarized in Algorithm \ref{algo:sol}. The \emph{status} of the final solution is either \emph{global} if the solution of LB-MIQP or LB-MIQCP satisfies Constraints \eqref{eq:sbounds}; \emph{infeas.} if one of the two relaxations returns infeasible; \emph{local} if UB-NLP returns a feasible upper-bound; or \emph{nosol.} if problem UB-NLP is infeasible. 

\begin{algorithm}
\small
\KwIn{$A$, $\bm{\theta}_0$, $\bm{v}_0$, $\qmin$, $\qmax$, $\tmin$, $\tmax$}
\KwOut{$\bm{q}^\star$, $\bm{\theta}^\star$, status}
$P \gets \{i \in A, j \in A : i<j\}$\\
$\bm{q}^\star, \bm{\theta}^\star, \bm{z}^\star \gets$ Solve LB-MIQP \\
\If{status(LB-MIQP)=infeas.}{
		status $\gets$ infeas.\\
		\Return
}
\If{$\bm{q}^\star \notin [\qmin, \qmax]$}{		
		$\bm{q}^\star, \bm{\theta}^\star, \bm{z}^\star \gets$ Solve LB-MIQCP \\
		\If{status(LB-MIQP)=infeas.}{
				status $\gets$ infeas.\\
				\Return
		}		
		\If{$\bm{q}^\star \notin [\qmin, \qmax]$}{
				status(LB-MIQCP) $\gets$ viol.\\
				$\bm{z} \gets \bm{z}^\star$\\
				$\bm{q}^\star, \bm{\theta}^\star \gets$ Solve UB-NLP\\
				\If{UB-NLP is feasible}{
						status $\gets$ local\\
				}
				\Else{
						status$\gets$ nosol.\\
				}
		}
		\Else{
			status$\gets$ global\\
		}
}
\Else{
	status $\gets$ global\\
}
\caption{Solution algorithm for the horizontal aircraft conflict resolution problem}
\label{algo:sol}
\end{algorithm} 
\normalsize

\section{NUMERICAL RESULTS}

\begin{figure}%
\begin{tabular}{ll}
\includegraphics[width=0.4\columnwidth]{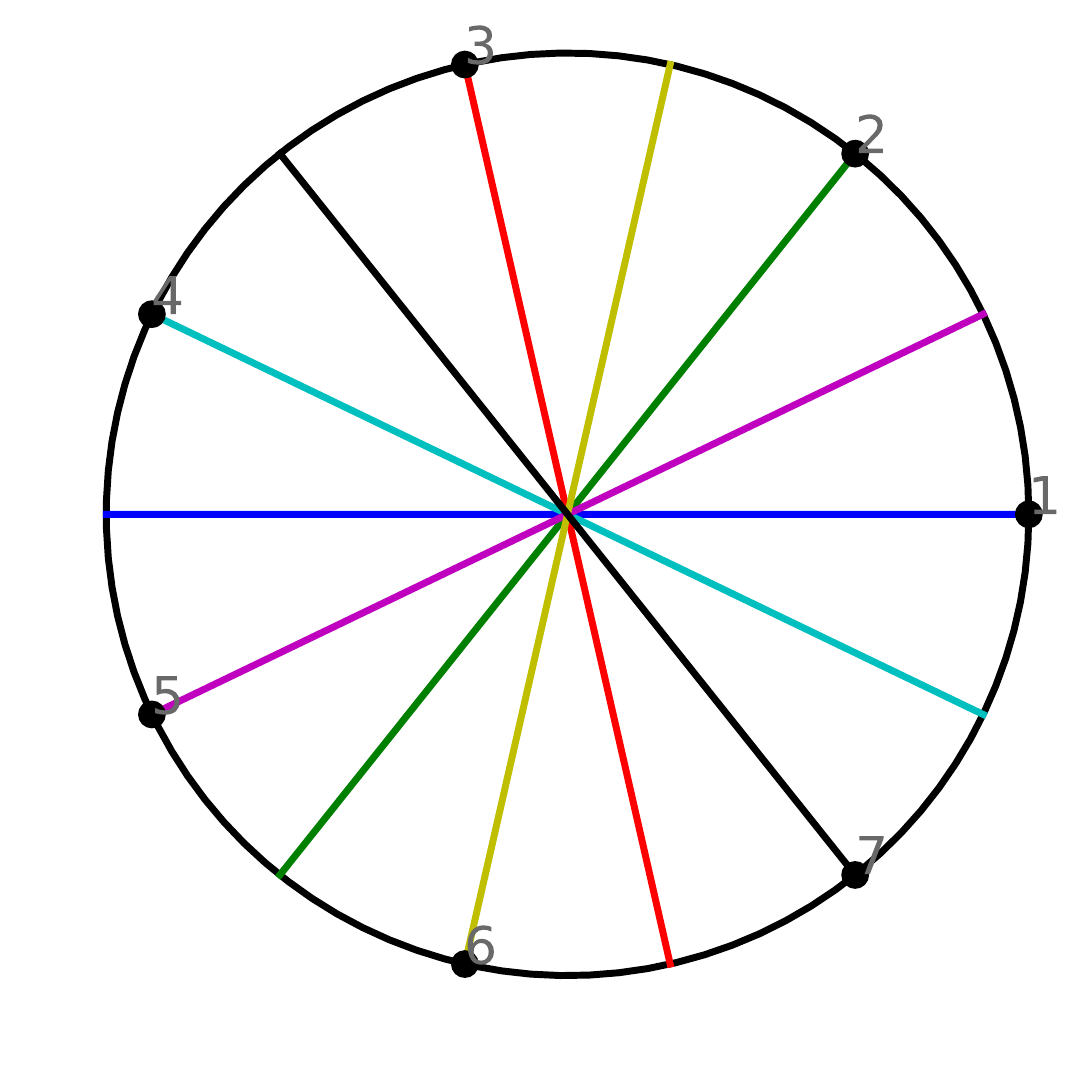} & \includegraphics[width=0.4\columnwidth]{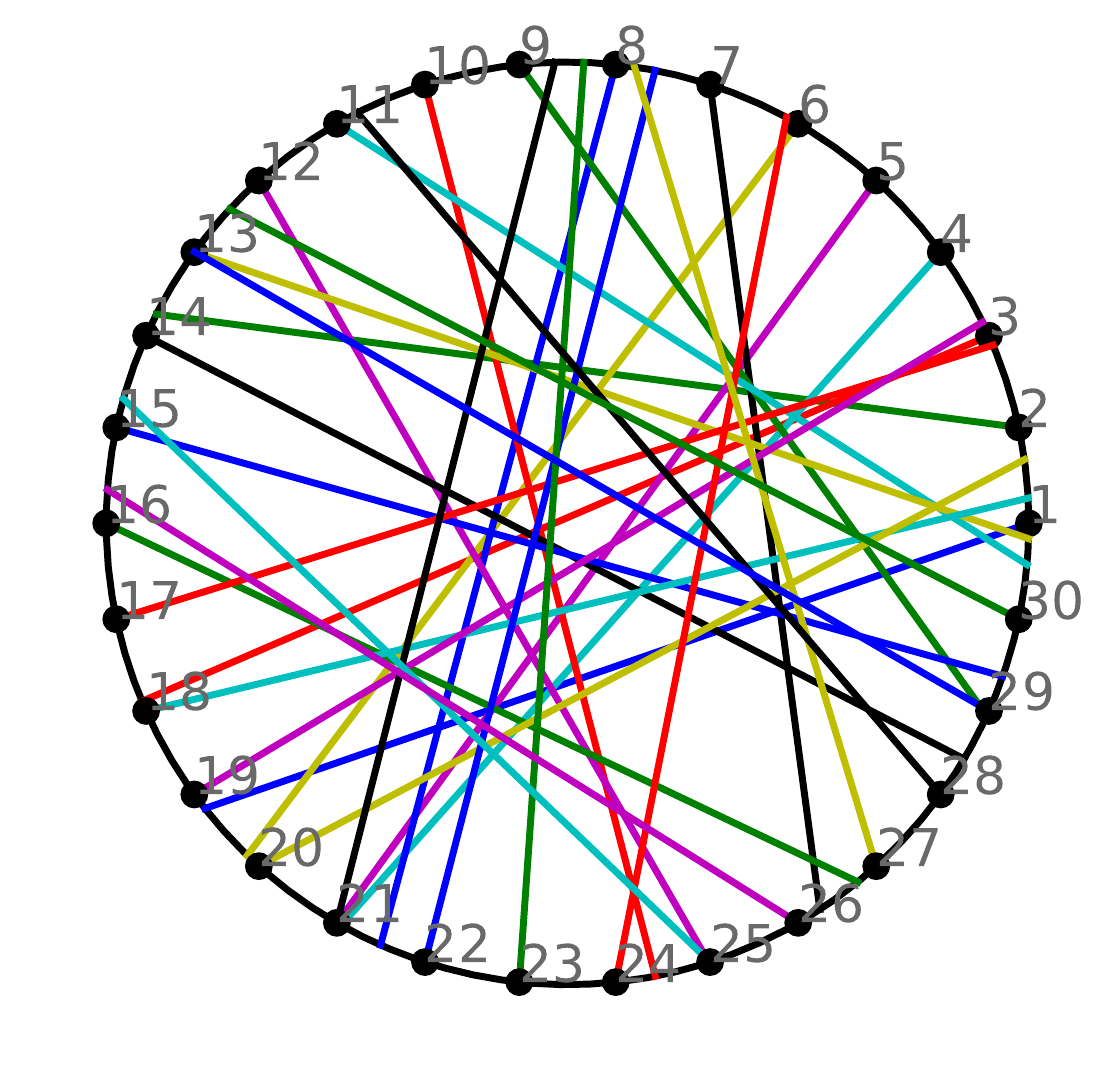}%
\end{tabular}
\caption{Illustration of benchmark instances: the CP with 7 aircraft (left) and the RCP with 30 aircraft (right).}%
\label{fig:bench}%
\end{figure}

We test the performance of the proposed complex number formulation with classical benchmark instances: the Circle Problem (CP) and the Random Circle Problem (RCP). The CP consists of a set of aircraft uniformly positioned on the circumference of a circle and heading towards its centre. Aircraft speeds are assumed to identical, hence the problem is highly symmetric. In contrast, the RCP builds on the same framework but aircraft initial speeds and headings are randomly deviated within specified ranges to create random instances with less structure. These benchmarks problems are illustrated in Figure \ref{fig:bench} and have been widely used in the field to assess the performance of CD\&R algorithms \cite{RRDW14,RRFF15,AEM16,CR17}. For reproducibility concerns, and for future comparisons, we have uploaded the models and the instances used here in the public repository \small\url{https://github.com/ReyHijazi/Conflict_Resolution}\normalsize.

In all experiments, we use a circle of radius of 200NM. For CP instances, all aircraft have the same initial speed of 500NM/h. For RCP instances, aircraft initial speeds are randomly chosen in the range 486-594NM/h and their initial headings are deviated from a radial trajectory (\emph{i.e.} towards the centre of the circle) by adding a randomly chosen an angle between $-\pi/6$ and $+\pi/6$. 

All considered models are implemented using the \Ampl modeling language~\cite{AMPL02} on personal computer with 8Gb of RAM and an Intel i7 processor at 2.9GHz. The MIQP and MIQCP problems are solved with \Cplex v12.7~\cite{CPLEX09} using default options and a time limit of 300s. The NLP problems are solved with \Ipopt with a constraint violation tolerance of $1e-9$~\cite{WB06}.

\subsection{Experimental Results in the Literature}
State-of-the-art computational results for the horizontal aircraft conflict resolution problem can be found in \cite{PFB02,Ome15,AEM16,CO16,LOSD17}. Some of these approaches \cite{PFB02,CO16} consider heading and speed control separately or maneuver-discretised formulations \cite{LOSD17} which can lead to suboptimal solutions. In \cite{AEM16}, an exact implementation of the combined speed and heading control problem is tested. Results on CP instances with more than 7 aircraft are not reported due to scalability issues. Asymmetric instances with up to 20 aircraft and an average number of conflicts of 18.6 are solved in 25-35s. 

\subsection{Our Experimental Results}

\begin{table*}%
\resizebox{1.9\columnwidth}{!}{%
\begin{tabular}{ll lllll lllll llll}
\toprule
&& \multicolumn{5}{l}{LB-MIQP} & \multicolumn{5}{l}{LB-MIQCP} & \multicolumn{4}{l}{UB-NLP}\\
\cmidrule(l){3-7} \cmidrule(l){8-12} \cmidrule(l){13-16}
$|A|$ & $n_c$ & Obj. & Time (s) & Gap (\%) & Status & $n_v$ & Obj. & Time (s) & Gap (\%) & Status & $n_v$ & Obj. & Time (s) & Gap (\%) & Status \\
\midrule
4	& 6		& 0.001250	& 0.453		& 0.000		& global	& -	& -	& -		& -		& -			& -	& -	& -	& -		& -			\\
5	& 10	& 0.002273	& 0.032		& 0.000		& global	& -	& -	& -		& -		& -			& -	& -	& -	& -		& -			\\
6	& 15	& 0.003619	& 0.062		& 0.022		& global	& -	& -	& -		& -		& -			& -	& -	& -	& -		& -			\\
7	& 21	& 0.004747	& 0.296		& 0.000		& global	& -	& -	& -		& -		& -			& -	& -	& -	& -		& -			\\
8	& 28	& 0.006921	& 2.199		& 0.009		& global	& -	& -	& -		& -		& -			& -	& -	& -	& -		& -			\\
9	& 36	& 0.008622	& 9.219		& 0.012		& global	& -	& -	& -		& -		& -			& -	& -	& -	& -		& -			\\
10	& 45	& 0.011099	& 73.367	& 0.010		& global	& -	& -	& -		& -		& -			& -	& -	& -	& -		& -			\\
11	& 55	& 0.013777	& 300.614	& 16.628	& local		& -	& -	& -		& -		& -			& -	& -	& -	& -		& -			\\
12	& 66	& 0.017111	& 301.098	& 44.899	& local		& -	& -	& -		& -		& -			& -	& -	& -	& -		& -			\\
13	& 78	& 0.019675	& 301.019	& 55.407	& local		& -	& -	& -		& -		& -			& -	& -	& -	& -		& -			\\
14	& 91	& 0.023641	& 301.066	& 67.744	& local		& -	& -	& -		& -		& -			& -	& -	& -	& -		& -			\\
15	& 105	& 0.028100	& 300.910	& 75.192	& local		& -	& -	& -		& -		& -			& -	& -	& -	& -		& -			\\
16	& 120	& 0.032525	& 300.863	& 80.290	& local		& -	& -	& -		& -		& -			& -	& -	& -	& -		& -			\\
17	& 136	& 0.037907	& 300.832	& 83.590	& local		& -	& -	& -		& -		& -			& -	& -	& -	& -		& -			\\
18	& 153	& 0.046422	& 300.770	& 87.040	& infeas.	& 2	& 0.046677	& 300.754	& 92.603	& local		& -	& -	& -	& -		& -			\\
19	& 171	& 0.055389	& 300.692	& 88.313	& infeas.	& 2	& 0.055572	& 300.677	& 93.593	& infeas.	& 2	& 0.057520	& 0.281	& 88.746	& local  \\
20	& 190	& 0.062233	& 300.536	& 91.056	& infeas.	& 4	& 0.062603	& 300.629	& 96.677	& infeas. & 1	& 0.064564	& 0.171	& 91.379	& local  \\
\bottomrule
\end{tabular}
}
\caption{Results on the Circle Problem.}
\label{tab:CP}
\end{table*}

We test the performance of Algorithm \ref{algo:sol} on the 17 CP instances with 4 to 20 aircraft. The results for these CP instances are summarized in Table \ref{tab:CP}. In the header, $|A|$ indicates the number of aircraft and $n_c$ indicates the number of initial conflicts $n_c$ (corresponding to the number of conflicts occurring if no control action is taken). The remaining of the table is organized in three sub-sections corresponding to the three steps of Algorithm \ref{algo:sol}. In each sub-section, \emph{Obj.} indicates the objective value, \emph{Time} indicates the computing time in seconds, \emph{Gap} indicates the relative optimality gap in \% and \emph{Status} indicates the status returned by Algorithm \ref{algo:sol}. Further, in subsections LB-MIQP and LB-MIQCP, $n_v$ indicates the number of bound-violating speed constraints \eqref{eq:sbounds}. A time limit of 300s is imposed for problems LB-MIQP and LB-MIQCP; and the status of the solution is set to \emph{local} if the solution returned after running out of time is feasible. In sub-section UB-NLP, the relative optimality gap is determined using the best of the lower bounds of LB-MIQP and LB-MIQCP. 

All CP instances with 4 to 10 aircraft are solved to global optimality within the allocating computing time. Instances with up to 7 aircraft are solved in less than a second. Instances with 11 to 17 are solved to local optimality within the first step of the algorithm (LB-MIQP) whereas instances 18 and 19 are solved in two steps (LB-MIQCP) and instance 20 is solved in three steps (UB-NLP). This is a substantial improvement compared to the existing literature where only results with up to 7 aircraft were reported \cite{AEM16}. 

\begin{table*}%
\begin{center}
\resizebox{1.3\columnwidth}{!}{%
\begin{tabular}{ll lllllll}
\toprule
&& \multicolumn{7}{l}{LB-MIQP} \\
\cmidrule(l){3-9} 
&  &  &  &  & \multicolumn{3}{l}{Status} &  \\
\cmidrule(l){6-8}
$|A|$ & $n_c$ & Obj. & Time & Gap (\%) & global & local & infeas. & $n_v$ \\
\midrule
10 & 3.1 (1.6) & 0.000444 (0.000) & 0.048 (0.013) & 0.003 (0.020) & 100 & 0 & 0 & 0.0 (0.0)  \\
20 & 13.1 (3.5) & 0.003540 (0.002) & 0.236 (0.077) & 0.005 (0.016) & 100 & 0 & 0 & 0.0 (0.0)  \\
30 & 32.9 (5.6) & 0.014369 (0.005) & 4.349 (4.359) & 0.003 (0.003) & 71 & 0 & 29 & 0.4 (0.6)  \\
40 & 59.3 (7.1) & 0.036929 (0.012) & 99.05 (88.99) & 0.555 (1.815) & 16 & 0 & 84 & 2.0 (1.6)  \\
\midrule
\\
&& \multicolumn{7}{l}{LB-MIQCP} \\
\cmidrule(l){3-9} 
&  &  &  &  & \multicolumn{3}{l}{Status} &  \\
\cmidrule(l){6-8}
$|A|$ & $n_c$ & Obj. & Time & Gap (\%) & global & local & infeas. & $n_v$ \\
\midrule
10 & 3.1 (1.6) & - (-) & - (-) & - (-) & - & - & - & - (-) \\
20 & 13.1 (3.5) & - (-) & - (-) & - (-) & - & - & - & - (-) \\
30 & 32.9 (5.6) & 0.019101 (0.003) & 35.58 (62.86) & 0.031 (0.119) & 12 & 0 & 17 & 1.0 (0.2) \\
40 & 59.3 (7.1) & 0.039747 (0.010) & 261.6 (73.23) & 9.655 (11.20) & 1 & 8 & 75 & 1.7 (0.8) \\
\midrule
\\
&& \multicolumn{7}{l}{UB-NLP}\\
\cmidrule(l){3-9} 
&  &  &  &  & \multicolumn{2}{l}{Status} & & \\
\cmidrule(l){6-7}
$|A|$ & $n_c$ & Obj. & Time & Gap (\%) & local & nosol. & & \\
\midrule
10 & 3.1 (1.6) & - (-) & - (-) & - (-) & - & - && \\
20 & 13.1 (3.5) & - (-) & - (-) & - (-) & - & - && \\
30 & 32.9 (5.6) & 0.021101 (0.004) & 0.249 (0.015) & 4.797 (8.819) & 17 & 0 && \\
40 & 59.3 (7.1) & 0.050284 (0.029) & 0.469 (0.051) & 13.85 (16.14) & 67 & 8 && \\
\bottomrule
\end{tabular}
}
\end{center}
\caption{Results on the Random Circle Problem.}
\label{tab:RCP}
\end{table*}

To evaluate the performance of the proposed approach on RCP instances, we generated 100 instances for four aircraft set sizes \emph{i.e.} 10, 20, 30 and 40 aircraft. The results are presented by reporting, for each aircraft set size, the mean over the 100 instances and the standard deviation in parenthesis. Solution status is reported by indicating the distribution of the 100 instances in whole numbers. 

All 200 10- and 20-aircraft RCP instances are solved to global optimality in one step, \emph{i.e.} after solving LB-MIQP, in less than a second. The initial conflict density in 30-aircraft RCP instances is more than twice that of 20-aircraft: 71 of them are solved to global optimality in one step while an additional 12 are solved to global optimality in two steps. All of the remaining 17 instances are solved to local optimality using the proposed heuristic with an average total computing time of 40s and an average optimality gap of 4.8\%. 40-aircraft instances pose a greater challenge with an average conflict density of 59.3. This leads to an average of 2 bound-violating constraints for LB-MIQP and 1.7 for LB-MIQCP. Consequently, only 17 of the 40-aircraft are solved to global optimality, while 75 are solved to local optimality and 8 out of 100 remain open. Most feasible solutions (bound-satisfying) are found using the heuristic with an average optimality gap of 13.9\% and a standard deviation of 16.1\%.

\section{CONCLUSIONS AND FUTURE WORKS}

We have introduced a novel formulation for the aircraft conflict resolution problem based on a complex number representation of velocity (speed and heading) control. The new model captures the non-convexity of the feasible region in a set of quadratic and linear on/off constraints. The resulting complex number formulation contains a single disjunction which models the crossing order of aircraft pairs at the intersection point of their trajectories. We introduce convex relaxations for this formulation and present a 3-step solution algorithm for its implementation. The performance of the proposed approach is tested on benchmark problems for conflict resolution. We find that the complex number formulation outperforms existing approaches and is able to solve to global optimality several open instances. Future work will be focused on multi-action control formulations to enable aircraft to return to their initial trajectories.

\bibliographystyle{IEEEtran}
\bibliography{IEEEabrv,CDR}

\begin{thebibliography}{10}
\providecommand{\url}[1]{#1}
\csname url@rmstyle\endcsname
\providecommand{\newblock}{\relax}
\providecommand{\bibinfo}[2]{#2}
\providecommand\BIBentrySTDinterwordspacing{\spaceskip=0pt\relax}
\providecommand\BIBentryALTinterwordstretchfactor{4}
\providecommand\BIBentryALTinterwordspacing{\spaceskip=\fontdimen2\font plus
\BIBentryALTinterwordstretchfactor\fontdimen3\font minus
  \fontdimen4\font\relax}
\providecommand\BIBforeignlanguage[2]{{%
\expandafter\ifx\csname l@#1\endcsname\relax
\typeout{** WARNING: IEEEtran.bst: No hyphenation pattern has been}%
\typeout{** loaded for the language `#1'. Using the pattern for}%
\typeout{** the default language instead.}%
\else
\language=\csname l@#1\endcsname
\fi
#2}}

\bibitem{Nol10}
M.~Nolan, \emph{Fundamentals of Air Traffic Control}.\hskip 1em plus 0.5em
  minus 0.4em\relax Cengage Learning, 2010.

\bibitem{ICA96}
ICAO, ``Rules of the air and air traffic services,'' International Civil
  Aviation Organization, Tech. Rep., 1996.

\bibitem{KY00}
J.~K. Kuchar and L.~C. Yang, ``A review of conflict detection and resolution
  modeling methods,'' \emph{Intelligent Transportation Systems, IEEE
  Transactions on}, vol.~1, no.~4, pp. 179--189, 2000.

\bibitem{RH02}
A.~Richards and J.~P. How, ``Aircraft trajectory planning with collision
  avoidance using mixed integer linear programming,'' in \emph{American Control
  Conference, 2002. Proceedings of the 2002}, vol.~3.\hskip 1em plus 0.5em
  minus 0.4em\relax IEEE, 2002, pp. 1936--1941.

\bibitem{PFB02}
L.~Pallottino, E.~M. Feron, and A.~Bicchi, ``Conflict resolution problems for
  air traffic management systems solved with mixed integer programming,''
  \emph{IEEE transactions on intelligent transportation systems}, vol.~3,
  no.~1, pp. 3--11, 2002.

\bibitem{VSSC09}
A.~Vela, S.~Solak, W.~Singhose, and J.-P. Clarke, ``A mixed integer program for
  flight-level assignment and speed control for conflict resolution,'' in
  \emph{Decision and Control, 2009 held jointly with the 2009 28th Chinese
  Control Conference. CDC/CCC 2009. Proceedings of the 48th IEEE Conference
  on}, Dec 2009, pp. 5219--5226.

\bibitem{RRFF15}
D.~Rey, C.~Rapine, R.~Fondacci, and N.-E. El~Faouzi, ``Subliminal speed control
  in air traffic management: Optimization and simulation,''
  \emph{Transportation Science}, vol.~50, no.~1, pp. 240--262, 2015.

\bibitem{Ome15}
J.~Omer, ``A space-discretized mixed-integer linear model for air-conflict
  resolution with speed and heading maneuvers,'' \emph{Computers \& Operations
  Research}, vol.~58, pp. 75--86, 2015.

\bibitem{AEM11}
A.~Alonso-Ayuso, L.~Escudero, and F.~Mart{\'{\i}}n-Campo, ``Collision avoidance
  in air traffic management: A mixed-integer linear optimization approach,''
  \emph{Intelligent Transportation Systems, IEEE Transactions on}, vol.~12,
  no.~1, pp. 47--57, March 2011.

\bibitem{AEM14}
A.~Alonso-Ayuso, L.~F. Escudero, and F.~J. Mart{\'\i}n-Campo, ``Exact and
  approximate solving of the aircraft collision resolution problem via turn
  changes,'' \emph{Transportation Science}, vol.~50, no.~1, pp. 263--274, 2014.

\bibitem{AEM16}
------, ``An exact multi-objective mixed integer nonlinear optimization
  approach for aircraft conflict resolution,'' \emph{TOP}, vol.~24, no.~2, pp.
  381--408, 2016.

\bibitem{CR17}
S.~Cafieri and D.~Rey, ``Maximizing the number of conflict-free aircraft using
  mixed-integer nonlinear programming,'' \emph{Computers \& Operations
  Research}, vol.~80, pp. 147--158, 2017.

\bibitem{CO16}
S.~Cafieri and R.~Omheni, ``Mixed-integer nonlinear programming for aircraft
  conflict avoidance by sequentially applying velocity and heading angle
  changes,'' \emph{European Journal of Operational Research}, 2016.

\bibitem{Caf14}
\BIBentryALTinterwordspacing
S.~Cafieri, ``{Maximizing the number of solved aircraft conflicts through
  velocity regulation},'' in \emph{{MAGO 2014, 12th Global Optimization
  Workshop}}, M{\'a}laga, Spain, Sept. 2014, pp. pp 1--4. [Online]. Available:
  \url{https://hal-enac.archives-ouvertes.fr/hal-01018051}
\BIBentrySTDinterwordspacing

\bibitem{Hij_MPC_16}
\BIBentryALTinterwordspacing
H.~Hijazi, C.~Coffrin, and P.~V. Hentenryck, ``Convex quadratic relaxations for
  mixed-integer nonlinear programs in power systems,'' \emph{Mathematical
  Programming Computation}, pp. 1--47, 2016. [Online]. Available:
  \url{http://dx.doi.org/10.1007/s12532-016-0112-z}
\BIBentrySTDinterwordspacing

\bibitem{Coff_16}
C.~Coffrin, H.~L. Hijazi, and P.~V. Hentenryck, ``The {QC} relaxation: A
  theoretical and computational study on optimal power flow,'' \emph{IEEE
  Transactions on Power Systems}, vol.~31, no.~4, pp. 3008--3018, July 2016.

\bibitem{Hij_10}
H.~Hijazi, ``{Mixed Integer NonLinear Optimization approaches for Network
  Design in Telecommunications},'' 2010, ph.D. thesis.

\bibitem{Hij_COA_12}
H.~Hijazi, P.~Bonami, G.~Cornu{\'e}jols, and A.~Ouorou,
  ``\BIBforeignlanguage{English}{Mixed-integer nonlinear programs featuring
  {"on/off"} constraints},'' \emph{\BIBforeignlanguage{English}{Computational
  Optimization and Applications}}, vol.~52, no.~2, pp. 537--558, 2012.

\bibitem{Hij_ANU_14}
H.~L. Hijazi, P.~Bonami, and A.~Ouorou, ``A note on linear on/off
  constraints,'' \emph{Australian National University technical report}, 2014.

\bibitem{CPLEX09}
I.~I. CPLEX, ``V12. 1: User’s manual for cplex,'' \emph{International
  Business Machines Corporation}, vol.~46, no.~53, p. 157, 2009.

\bibitem{RRDW14}
D.~Rey, C.~Rapine, V.~V. Dixit, and S.~T. Waller, ``Equity-oriented aircraft
  collision avoidance model,'' \emph{IEEE Transactions on Intelligent
  Transportation Systems}, vol.~16, no.~1, pp. 172--183, 2015.

\bibitem{AMPL02}
R.~Fourer, D.~M. Gay, and B.~W. Kernighan, \emph{{AMPL: A Modeling Language for
  Mathematical Programming}}, 2nd~ed.\hskip 1em plus 0.5em minus 0.4em\relax
  Brooks/Cole, 2002.

\bibitem{WB06}
A.~W{\"a}chter and L.~T. Biegler, ``On the implementation of an interior-point
  filter line-search algorithm for large-scale nonlinear programming,''
  \emph{Mathematical programming}, vol. 106, no.~1, pp. 25--57, 2006.

\bibitem{LOSD17}
T.~Lehouillier, J.~Omer, F.~Soumis, and G.~Desaulniers, ``Two decomposition
  algorithms for solving a minimum weight maximum clique model for the air
  conflict resolution problem,'' \emph{European Journal of Operational
  Research}, vol. 256, no.~3, pp. 696--712, 2017.

\end{thebibliography}

\end{document}